\documentclass[sigconf,screen]{acmart}

\AtBeginDocument{%
  \providecommand\BibTeX{{%
    \normalfont B\kern-0.5em{\scshape i\kern-0.25em b}\kern-0.8em\TeX}}}

\copyrightyear{2025}
\acmYear{2025}
\setcopyright{cc}
\setcctype[4.0]{by}

\acmConference[PETRA '25]{The PErvasive Technologies Related to Assistive Environments}{June 25--27, 2025}{Corfu Island, Greece}
\acmBooktitle{The PErvasive Technologies Related to Assistive Environments (PETRA '25), June 25--27, 2025, Corfu Island, Greece}
\acmDOI{10.1145/3733155.3736593}
\acmISBN{979-8-4007-1402-3/25/06}

\usepackage{tabularx}
\usepackage[capitalise,noabbrev]{cleveref}

\begin{document}

\title{Perceptions of Blind Adults on Non-Visual Mobile Text Entry}

\author{Dylan Gaines}
\orcid{0000-0002-2747-7680}
\affiliation{%
  \institution{Michigan Technological University}
  \department{Department of Computer Science}
  \city{Houghton}
  \state{Michigan}
  \country{USA}
}
\email{dcgaines@mtu.edu}

\author{Keith Vertanen}
\orcid{0000-0002-7814-2450}
\affiliation{%
  \institution{Michigan Technological University}
  \department{Department of Computer Science}
  \city{Houghton}
  \state{Michigan}
  \country{USA}
}
\email{vertanen@mtu.edu}

\renewcommand{\shortauthors}{Gaines and Vertanen}

\renewcommand{\shorttitle}{Perceptions of Blind Adults on Non-Visual Mobile Text Entry}

\begin{abstract}
Text input on mobile devices without physical keys can be challenging for people who are blind or low-vision. We interview 12 blind adults about their experiences with current mobile text input to provide insights into what sorts of interface improvements may be the most beneficial. We identify three primary themes that were experiences or opinions shared by participants: the poor accuracy of dictation, difficulty entering text in noisy environments, and difficulty correcting errors in entered text. We also discuss an experimental non-visual text input method with each participant to solicit opinions on the method and probe their willingness to learn a novel method. We find that the largest concern was the time required to learn a new technique. We find that the majority of our participants do not use word predictions while typing but instead find it faster to finish typing words manually. Finally, we distill five future directions for non-visual text input: improved dictation, less reliance on or improved audio feedback, improved error correction, reducing the barrier to entry for new methods, and more fluid non-visual word predictions.
\end{abstract}



\maketitle

\section{Introduction}
For many adults, text input is an extremely common daily task. Often this text input is on a mobile device without a physical keyboard. Without the physical boundaries between keys providing tactile feedback as to the key locations, and without being able to feel the key actuations to know when they are selected, mobile text input can be difficult for people who are blind or low-vision (BLV).
When researching any Human-Computer Interaction (HCI) topic, it is imperative to consider the specific needs of the target users. When touchscreen smartphones first became commercially popularized, many studies were conducted that interviewed people who were BLV. Since then, there has been a great deal of research on non-visual text input methods and many commercial solutions, but few recent interview studies on how people who are BLV perform mobile text input on a day-to-day basis. To our knowledge, only one interview study has been published regarding non-visual text input in the last five years. This study by \citet{karimi-nomadic} focuses on a specific context (text input while traveling) as opposed to the general case of non-visual text input.

As technology changes, so too can the needs of the users. In this work, we aim to assess the current user needs in non-visual text input and provide recommendations for future research. We interview 12 legally blind adults about their past and present experiences with mobile text input and gather their thoughts on different interaction techniques, including an experimental research prototype.

\section{Related Work}
\subsection{Braille-Based Text Input}
A wide variety of interfaces based on the Braille alphabet have been developed to try to improve non-visual text input. Perkinput~\cite{azenkot-perkinput} encoded characters as six-bit binary strings using characters' Braille representation. These binary strings were entered using either three fingers on each hand simultaneously or using two sequential three-fingered touches with a single hand. Perkinput users typed on average at 17.6 words per minute with an uncorrected error rate of 0.14\% using a single hand, or 38.0 words per minute with an error rate of 0.26\% using two hands.

In BrailleTouch~\cite{southern-brailletouch}, a user holds the device with the screen facing away from them and uses the first three fingers on each hand to tap the Braille encoding for characters. Expert Braille typists obtained an average entry rate of 23.2 words per minute with a 14.5\% error rate in their final of five sessions using BrailleTouch on a smartphone. For comparison, they averaged 42.6 words per minute with a 5.3\% error rate in the final session using a physical Braille keyboard.

In contrast to Perkinput~\cite{azenkot-perkinput} and BrailleTouch~\cite{southern-brailletouch} which split the 3\,$\times$\,2 Braille matrix into left and right sides, TypeInBraille~\cite{mascetti-typeinbraille} allowed users to enter characters one row at a time (i.e. using three actions instead of two). Users tapped on the left or right side (or both) to indicate which dots were raised, or with three fingers to indicate no dots in that row. Swiping to the right indicated the end of a character. TypeInBraille users typed on average at around 7 words per minute with just under a 5\% error rate.

BrailleType~\cite{oliveira-brailletype} placed six targets on a touchscreen corresponding to the dot locations in a Braille character, one in each corner and one along each of the two long edges. Users marked dots by dragging their finger to each dot and waited for an audio confirmation at each location. Double tapping on the screen input the character represented by the currently marked dots, and swiping to the left cleared any marked dots or deleted the last character if no dots were marked. BrailleType users entered text at 1.45 words per minute with an 8.91\% error rate.

Commercial solutions have been developed with similar techniques to these research systems. The MBraille keyboard\footnote{\url{https://mpaja.com/mbraille/en.lproj/help.htm}} allows users to decide where to place each dot on the screen. Users enter characters by touching each dot in a character's encoding at the same time, and the character is typed when the fingers are released. Alternate modes allow for input with the screen facing away from the user, similar to BrailleTouch~\cite{southern-brailletouch}, and with all dots arranged horizontally similar to the bimanual entry mode in Perkinput~\cite{azenkot-perkinput}.

The VoiceOver accessibility software developed by Apple and built into the iOS operating system also allows for Braille-based text input with its Braille Screen Input (BSI).\footnote{\url{https://support.apple.com/guide/iphone/type-onscreen-braille-iph10366cc30/ios}} Similar to BrailleTouch~\cite{southern-brailletouch}, BSI also contains a screen-away mode that allows users to reposition the Braille dots.

\subsection{Other Non-Visual Text Input}
Other research has focused on ways to make the typical Qwerty keyboard more accessible to people who are BLV. The item selection technique used by Slide Rule~\cite{kane-slide} allows users to scan through a list of items by swiping their finger across the screen and listening to audio feedback. A user can select the item they are currently on by tapping a second finger in a different location. While the studies performed in this paper did not focus on text input, the authors noted that their system enabled text input using a keyboard. 

A similar technique can be used with an onscreen keyboard using Apple's VoiceOver.\footnote{\url{https://support.apple.com/guide/iphone/use-the-onscreen-keyboard-iph3e2e3d1d/ios}} In addition to allowing selection by tapping a second finger (split-tapping), users can perform a double-tap gesture to enter a selected key or simply enter the selected key when they lift their finger. A VoiceOver setting allows users to choose their desired selection method. 

On many devices, users can also dictate their text and have it be entered by a speech recognition algorithm. This dictation, or speech input, serves as an alternative to typing altogether. Speech input is available, for example, on Apple's iOS keyboard\footnote{\url{https://support.apple.com/en-ie/guide/iphone/iph2c0651d2/ios}} and Google's Gboard\footnote{\url{https://play.google.com/store/apps/details?id=com.google.android.inputmethod.latin}} (available for both iOS and Android operating systems).

The commercial FlickType keyboard\footnote{\url{https://www.flicktype.com/}} allows users to enter text by tapping approximate character locations instead of definitively locating each key. By swiping to the right, the user signals to the system that they are finished entering the word. The system produces a best guess of the user's intended word from their tap locations, and the user can swipe through the list of suggested words. 

Tinwala and MacKenzie developed a \emph{Graffiti}-based approach~\cite{tinwala-graffiti} in which users would perform gestures similar to drawing each character on the screen. They found that users entered text with at an average of 10.0 words per minute with an error rate of 4.3\%.

We previously developed FlexType~\cite{gaines-flextype}. FlexType allows users to enter text via ambiguous gestures by tapping anywhere on the screen with between one and four fingers at the same time to indicate one of four groups of characters. Swiping right at the end of each word signals the software to determine the most likely word that matches the tap sequence given the context of what the user has already written. We found users had entry rates around 12.0 words per minute with error rates around 2.0\%. In this work, we describe this text input method to our participants to solicit their thoughts and concerns on the technique. Portions of the interview presented here were previously described briefly in \cite{gaines-improving} as they relate to the further development of FlexType. This work provides additional detail and analyzes the results from a broader perspective to relate findings to non-visual text input in general as opposed to any specific interface.

\subsection{Interviews with People who are BLV}
Early work on accessible technology incorporated interviews and case studies to learn more about how people who were BLV interacted with technology~\cite{shinohara-designing}. The work done by~\citet{kane-slide} that led to the Slide Rule selection method focused on formative interviews that explored users' interactions with touchscreens.

Further work by~\citet{azenkot-exploring} sought to investigate how people who were blind used speech input on mobile devices compared to people who were sighted. They conducted a survey with 169 people (of which 64 were BLV) and found that people who were BLV were more satisfied with speech input and felt it was faster compared to people who were sighted. 

\citet{abdolrahmani-empirical} interviewed eight people who were BLV to learn how they use their devices in certain situations, such as when their hands are occupied or when they were on crowded public transportation. Their participants expressed concerns about privacy and discretion while using their mobile devices in public.

Recent work by~\citet{karimi-nomadic} conducted interviews with 20 people who were BLV and explored how they text while traveling. They found that texting on-the-go often required users to switch between applications in order for them to obtain information about their surroundings (e.g.~obstacles in their way). While this work sought to explore the different factors that impact text input while moving, we focus our interview on the general case for non-visual input, without any specific context of use.

\begin{table*}[tb]
  \begin{center}
    \begin{tabular}{ l r r r }
    \toprule
      Participant & Visual Acuity & Duration of Blindness & Cause of Bindness \\
    \midrule
    P1 & 0 & Legally 45 yr, completely 26 yr & Retinitis pigmentosa \\
    P2 & 20/3000, 5 degree visual field & About 20 yr & Retinitis pigmentosa \\
    P3 & 0 & Since birth & Microphlalmia \\
    P4 & 0 & Since birth & Retinopathy of prematurity \\
    P5 & 20/10000 & 19 yr & Leber hereditary optic neuropathy \\
    P6 & 0 in one eye, 20/400 in other & Since birth & Cataracts \\
    P7 & Minimal light perception & Since birth & Leber congenital amaurosis \\
    P8 & 0 in one eye, 20/400 in other & Since birth & Retinopathy of prematurity \\
    P9 & 0 & 42 yr & Retinal detachment \\
    P10 & 20/400 & Since birth & Leber congenital amaurosis \\
    P11 & 20/300 & Since birth & Brain cyst, nystagmus \\
    P12 & 0 & Since birth & Dark corneas, cataracts \\
    \bottomrule
    \end{tabular}
  \caption{Details of the visual acuity, duration of blindness, and cause of blindness for each participant.}
  \label{condition_table}
  \end{center}
\end{table*}

\section{Semi-Structured Interview}
We recruited a total of 12 legally blind adults for this study via the National Federation of the Blind mailing list in the United States. Participants were selected based on the order that they responded to our advertisement. The interviews took approximately 30--60 minutes and were conducted via Zoom. Participants received a US\$20 Amazon gift card as compensation.

\subsection{Methods}
At the beginning of each interview, we obtained oral consent from the participant and then asked a series of demographic questions about their age, gender, and blindness. We asked each participant about their typical use cases for text input on mobile devices, and which languages they entered text in. We then asked about the main text entry technique they used, and what they liked and did not like about it. We followed this by asking about any other techniques that they use now or have used in the past. We asked the participants about common sources of error in their entered text, and their experiences detecting and correcting those errors. 

To provoke discussion on various types of interfaces, we asked participants questions about their knowledge of Braille, as well as any experiences with Braille-based input on mobile devices. To probe participants' willingness to learn new text input methods, we then described our FlexType interface~\cite{gaines-flextype}. We described FlexType as follows:

``FlexType would remove all dependence on the location of your taps. Essentially, it divides the letters into four groups and you tap with the number of fingers corresponding to the group containing your letter. For example, group 1 contains A through E, so for any of those letters you would tap with one finger. Group 2 contains F through M, so for any of those you'd tap with two fingers. You'd tap with 3 fingers for N through R, and with 4 for S through Z and apostrophe. Once you do all of the taps for a word, you swipe to the right and the interface uses the sequence of taps as well as the context of what you have already written to determine a list of the most likely matching words. You would then swipe up and down to navigate through this list until you hear the word you were trying to type.''

We concluded the interview by discussing participants' experiences with word predictions, how their main interface currently presents predictions (if at all), the speaking rate setting they use for text-to-speech (TTS), and how frequently they use earbuds during text input. This portion of the interview was conducted to evaluate the efficacy of current word predictions and the potential of further research on audio word predictions.

To analyze the results of our interview, we first transcribed each audio recording. We then summarized each participants' response to each question or topic in a chart. From the completed chart, we identified commonalities among participants' opinions. We sought to identify experiences or opinions that many of our participants shared to help guide the direction of non-visual text input research. For each commonality, we reviewed each interview's transcription to gather the context of our participants' comments. We then developed a theme that we felt best reflected the overall shared view of our participants.

\begin{table*}[tb]
  \begin{center}
    \begin{tabular}{ l r r r }
    \toprule
      Input Method & Primary & Secondary & Previously Used \\
    \midrule
    Dictation & P1, P2, P3, P6, P7, P9, P11 & P4, P5, P10 & - \\
    Onscreen keyboard with VoiceOver & P4, P5, P10, P12 & P1, P3, P6, P7, P9, P11 & - \\
    Onscreen keyboard without VoiceOver & - & P8 & - \\
    Wireless physical keyboard & - & P5, P8, P9 & P6, P12 \\
    Braille Screen Input (BSI) & P8 & P4 & P12 \\
    MBraille & - & - & P4 \\
    FlickType (or similar method) & - & - & P10, P12 \\
    \bottomrule
    \end{tabular}
  \caption{The participants who reported using each input method as their primary, secondary, or previously used method.}
  \label{input_method_table}
  \end{center}
\end{table*}

\subsection{Demographics}
Participants ranged in age from 38 to 66 (mean 50.3). Six identified as female, five as male, and one did not identify with either gender. None of the participants were currently studying at a university. Five participants were completely blind, with one more having only minimal light perception. Two additional participants were completely blind in one eye with a 20/400 visual acuity in the other eye. Eight participants reported being blind since birth, and all participants had been blind for a minimum of 19 years. More details on the participants' visual acuities and causes of blindness can be found in \cref{condition_table}.

\subsection{Results}
\subsubsection{Types of Text}
When asked what types of text they typically wrote on mobile devices, all 12 participants reported sending text messages, 8 reported composing emails, and 7 reported performing web searches. Three participants said that they interacted with social media from their mobile devices, and another three filled out forms or surveys. One participant reported composing short stories or novels entirely on their mobile device. Five of the participants reported entering text in languages other than English, though four of them specified that this was rare.

\subsubsection{Input Methods}
\cref{input_method_table} shows which participants reported using each input method. Dictation was the most common primary text input method, with seven participants reporting using it as their primary way of entering text and another three using it as a secondary method. Many cited its speed as the main benefit of inputting text by voice. Four participants used the onscreen keyboard with VoiceOver as their primary method, and six more as a secondary method. The ways participants confirmed selections via VoiceOver was a mix between double-tapping, split-tapping, and type-on-release. One participant reported using the soft keyboard without a screen reader as a secondary method. Only one participant reported using Braille Screen Input as their primary text input method, and one other participant used it as a secondary method. A third participant reported having used it in the past, and another mentioned having used MBraille previously. Three participants used Bluetooth keyboards as a secondary input method (and two more had used one previously), and two others mentioned having used an interface similar to FlickType in the past.

\subsubsection{Identified Themes}
The first theme that we identified from our interviews was the \textbf{poor accuracy of dictation}. All ten participants that reported using dictation as either a primary or secondary text input method specifically mentioned that they had issues with the speech recognizer's ability to correctly determine what they said. This presented in a variety of ways to the different participants, some citing background noise, accents, uncommon words, or artifacts in their own speech such as coughs or hiccups as the reasons for recognition errors. Supporting quotes for each theme can be found in \cref{themes_table}.

This lead to the next theme that we identified: participants frequently mentioned text input was \textbf{difficult in noisy environments}. Users of both dictation and a soft keyboard with VoiceOver mentioned having these difficulties; dictation users mentioned that the speech recognizer often picked up on background noise, while soft keyboard users struggled to hear the audio feedback from their device. 

In public places such as restaurants or on public transit, another common theme that emerged was participants were \textbf{concerned about their privacy}. Participants were wary of other people hearing their dictated messages, or hearing the audio feedback from their devices. The participant that used Braille Screen Input primarily used screen away mode. They were frustrated that the text they were composing was unnecessarily displayed on the screen for anyone to see.

\begin{table*}[tb]
  \small
  \begin{center}
    \begin{tabularx}{\textwidth}{ l X }
    \toprule
      Theme & Supporting Quotes \\
    \midrule
    Poor accuracy of dictation & ``I used it to write a text message to my mom one time and the text came out all wrong. So I think that there's some issues with the microphone picking up the right words and things.'' (P11) \newline \newline
    ``It doesn't understand all of what I'm saying, and it writes the wrong word, and then I have to go back and edit it several times to make it say the right thing.'' (P7) \\
    \midrule
    Difficult in noisy environments & ``When I'm in a noisy environment I have the the same issue that dictation does, which is hearing VoiceOver speak. So a lot of the characters kind of blend into each other when I'm trying to type something in a noisy environment.'' (P1) \newline \newline
    ``If it's crowded, the dictation will hear other voices, and then it will get confused about what I'm saying.'' (P9) \\
    \midrule
    Concerned about their privacy & ``I have one ear bud so I can always be listening to stuff on my phone and I can interact with it, and nobody else has to hear my VoiceOver that way, because it's a privacy thing.'' (P5) \newline \newline
    ``One thing that drives me nuts about BSI is that there's no way to make it not show the words you're typing. I am a little bit creeped out that since I have my phone facing away from me, anybody can just read what I'm typing.'' (P8)\\
    \midrule
    Difficulty correcting errors  & ``I haven't figured out an easy way to navigate the text and edit it.'' (P2) \newline \newline
    ``It's pretty easy to detect an error when it reads it back to me. It's just it's harder to go back and correct it.'' (P3) \\
    \midrule
    Cumbersome to carry a wireless keyboard &  
    ``The thing that I find frustrating with that is like, I usually don't have those devices with me and connected when I wanna send, you know, like a text or something.'' (P12) \\
    \midrule
    Degradation in third party app usefulness & ``I really loved that app a lot, but then they changed it a bunch and it didn't work as well anymore.'' (P12) \\
    \midrule
    Steep learning curve for new methods & ``I feel like I would have trouble onboarding and it feels like for me, I think I could see myself getting really frustrated really fast.'' (P8) \newline \newline
    ``I'm generally not a fan of a text entry method that involves me having to sort of relearn text entry. That's always the thing that sort of kills it for me is that I have to now have to enter this new thought process into my head.'' (P10) \\
    \midrule
    Word predictions were disruptive & ``Those suggestions weren't just physically disrupting with typing, which they were, they were also really mentally disruptive.'' (P8) \newline \newline
    ``I can tell you I've never used it, and when it's come up by mistake, I've hated it and had to redo it. You know it's it's just been a complete hindrance to me.'' (P6) \\
    
    \bottomrule
    \end{tabularx}
  \caption{The themes we identified from our interviews with supporting participant quotes.}
  \label{themes_table}
  \end{center}
\end{table*}

Most participants noted that it was typically fairly easy to detect when an error had occurred via the interface reading their text back to them. However, participants frequently mentioned they had \textbf{difficulty correcting errors}. A few were unsure of how to move the cursor back in the text field to get to the error, opting instead to delete their text and start over or to send a follow-up message to their recipient correcting the error. Some participants elaborated that when moving the cursor through the text field, it was difficult to know whether the cursor was at the beginning or the end of the word spoken by the screen reader. 

Of the five participants that talked about using a Bluetooth wireless keyboard, three mentioned that the reason they did not use it more frequently was that while it is quite fast, it was \textbf{cumbersome to carry a wireless keyboard}. 

The three participants that discussed MBraille and FlickType mentioned that often times updates to their devices' operating systems or the application itself caused a \textbf{degradation in third party app usefulness}. Participants noted often third-party developers failed to offer long-term support for their methods. 

Participants had mixed enthusiasm about the FlexType interface~\cite{gaines-flextype}. Eight of the twelve were interested in at least trying it, but two were firmly against the idea. One participant did not take a firm position either way, and the final participant was hesitant at first, but concluded their interview by stating ``it sounds really promising'' (P8). The chief concern that participants voiced was the \textbf{steep learning curve for new methods} such as FlexType. In particular they were concerned with memorizing the groups of characters. Participants were also concerned about the accuracy of the algorithm in determining their intended word. One of the original limitations of FlexType~\cite{gaines-flextype} was that users were unable to enter words that did not appear in the top six predictions for that combination of group sequence and context. When asked how they would envision entering uncommon words or proper names that might not be in the system's vocabulary, the most common responses were to allow users to define a custom dictionary and to have a letter-at-a-time mode that would remove the ambiguity from characters. Based on this feedback, we implemented a letter-entry mode in the next iteration of FlexType~\cite{gaines-improving}.

When we asked participants about their usage of word predictions, seven participants either ignored them or had the option disabled altogether, one reported using them very rarely, and the other four said they actively used them. All four participants that actively used word predictions said that they were displayed visually on the screen and they needed to explicitly explore them with their screen reader and then select them like they would a key. Participants that did not use word predictions often said that the \textbf{word predictions were disruptive}, and that it was easier or faster for them to just finish typing the word. Of the four participants that reported having predictions enabled, one reported that they used them frequently, two reported using them about half the time, and one reported using them only when they were unsure about how to spell a word.

Text-to-speech (TTS) speeds used by our participants ranged from the default speed (denoted as 50\% in VoiceOver) to 95\% out of a maximum possible 100\%. Participant use of earbuds while entering text was also mixed, with some saying they would use them only when around other people and others using them the majority of the time. Two participants mentioned only using a single earbud at a time so that they were still able to hear their surroundings (e.g.~listening for their stop on public transit). Details on TTS speeds and earbud usage can be found in \cref{tts_table}.

\begin{table}[tb]
  \begin{center}
    \begin{tabular}{ l r r }
    \toprule
      Participant & TTS Speed & Earbud Usage \\
    \midrule
    P1 & Not reported & Hearing aid via Bluetooth \\
    P2 & Not reported & Not reported \\
    P3 & 70\% & When privacy is a concern \\
    P4 & 70--75\% & In the office, but not at home \\
    P5 & 95\% & 80\% of the time, single earbud \\
    P6 & Not reported & Not reported \\
    P7 & 85\% & 75\% of the time \\
    P8 & 90\% & 10\% of the time \\
    P9 & Not reported & None \\
    P10 & 75\% & 90\% of the time, single earbud \\
    P11 & Default (50\%) & When around other people \\
    P12 & 65\% & Nearly all the time \\
    
    \bottomrule
    \end{tabular}
  \caption{Participants' text-to-speech speed and earbud usage.}
  \label{tts_table}
  \end{center}
\end{table}

\section{Directions for Future Research}
\subsection{Improvements to Dictation}
Our interviews show that dictation is the preferred text input method of many people who are BLV, even in spite of the concerns they raised about its accuracy. This is due mainly to the speed at which users are able to enter text, even if they need to dictate their message multiple times to correct errors. We found it somewhat surprising that recognition accuracy was a pain point for users given all the progress in recognition accuracy resulting from the use of neural networks~\cite{hinton2012deep}. 

We recommend further work on the models behind the speech recognizer, specifically to make them more robust when the input contains background noise, accents, or artifacts such as coughs. It could also be useful to further develop interfaces that help the user avoid errors in the first place, for example by spelling difficult words in their input~\cite{vertanen-spelling}. By reducing the error rate of dictated text, we can enable its use in more circumstances and improve the efficiency at which people who are BLV can input text on mobile devices. 

Another area of dictation that could be improved is how it treats pauses in speech. P9 noted that ``If I slow down or stop, you know, trying to compose a sentence, it will try to send the message right away without me completing my thought or my sentence.'' It can be difficult for a user to know the full message that they wish to send before they begin typing it, especially in the case of longer messages or emails. When typing on a keyboard it is easy to pause to think, but further work is needed to support intermittent speech input for people who are BLV.

\subsection{Rely Less On or Improve Audio Feedback}
While dictation has room to improve significantly, an excess of background noise or privacy concerns could still call for the use of an alternative text input method. Both of these scenarios could benefit from a text input method that has less reliance on audio feedback. One participant suggested using vibration to convey information to the user in addition to, or in lieu of, audio feedback. While there has been some work on this in the form of a Braille-based wearable glove for people who are deaf-blind~\cite{choudhary-glove}, there is still an opportunity for further research in how vibration feedback can help non-visual text input. 

When audio feedback is needed, modifying the audio signal may be able to improve its intelligibility in noise. This is a long-standing and ongoing area in speech research, e.g.~the Hurricane Challenge~\cite{cooke_hurricane,rennies_hurricane2}. In the case of text input interfaces, specific research into the intelligibility of short audio segments such as individual letters or words is needed. It is also possible to direct the audio to a specific location in space via beamforming~\cite{watanabe2009a}. If a mobile phone supported beamformed output, this could allow targeting more signal amplitude to a user's head position (inferred by other sensors like a microphone array). This would have the added benefit of reducing its amplitude to nearby non-users, potentially reducing privacy concerns while avoiding the need for headphones that can restrict a user's ability to hear their surroundings.

\subsection{Improved Error Correction}
Errors are an inevitable part of text input, whether they originate from the system or from the user. Efficiently correcting errors is vital to achieving acceptable overall entry rates~\cite{rodrigues-open}. Recent work by~\citet{zhang-type} proposed three ways that sighted users could go about correcting errors without navigating the cursor to the location of an error. While two of these involved dragging a correction to the location of the error and would likely not be well-suited to non-visual text input, the Magic Key technique used a recurrent neural network to determine the most likely errors in a user's text. This method could be adapted to make non-visual error correcting more efficient. 

To correct errors in dictation, some work has explored allowing users to re-speak an erroneous section, instead of the entire utterance~\cite{mcnair-improving,vertanen-automatic,ghosh-commanding}. This serves to prevent recognition errors from occurring in different parts of the text on the second attempt, and may be able to detect where the correction needs to be applied without the user explicitly specifying the location. While these initial research studies have shown promise, such features have yet to be implemented in commercially available speech recognition systems. Further research in this area may help to increase its adoption.

\subsection{Reduce Barrier to Entry for New Methods}
While only two of the twelve participants were not interested in trying the FlexType interface~\cite{gaines-flextype}, many were hesitant to fully adopt it because they were concerned about the barrier to entry created by the novel technique. We recommend future research consider ways to reduce the barrier to entry for any new input method. One possibility for this is to maintain some elements of consistency with current popular input methods, so users do not feel they need to enter an entirely new thought process when entering text.

While not necessarily a non-visual input method, an example of this research direction is the SHARK$^2$ system~\cite{kristensson-shark2}. SHARK$^2$ created a technique where words could be written by tracing a shape through all the letters of a word on an onscreen keyboard instead of individually tapping each key. Originally created for pen-based interaction, this technique has developed into the word-gesture keyboard~\cite{zhai_reimagining}, a common feature on touchscreen keyboards. Instead of replacing the familiar text input method altogether, the word-gesture keyboard allows a user to fluidly switch between standard tapping and gesture input and to leverage their existing knowledge of the Qwerty keyboard layout.

\subsection{Fluid Non-Visual Word Predictions}
The text input interfaces our participants currently use typically provide word predictions in a visual manner by displaying them above the top row of keys. Non-visually, these predictions need to be explored via a screen reader. Many of our participants reported that these predictions were difficult to use non-visually and were even disruptive if selected accidentally. We believe that further research on non-visual word predictions could allow them to be beneficial to users instead of a hindrance. 

An example of research aligned with this direction is the use of simultaneous audio to present multiple word predictions at once~\cite{montague-inviscid,gaines-simultaneous}. This could avoid the need for a user to interrupt their typing process to receive word predictions, and increase the rate at which the predictions are presented. The design space of this topic was explored by~\citet{nicolau-design}, but further development and adoption of this technique in commercially available interfaces could have a widespread positive impact across users who are BLV.

Another possible interaction would be to incorporate predictions more fluidly in the typical text input workflow for Qwerty keyboards with VoiceOver. For example, if while typing a user pauses for longer on their next intended key, the system could begin to read the system's top word predictions. The user could then make a special gesture to select the prediction instead of typing just the key's letter.

\section{Discussion and Limitations}
The primary aim of this work was to gather information about the day-to-day use of current non-visual text input methods and assess the needs of users who are BLV. In the prior section, we recommended five directions for future research that our interviews suggest would have the biggest impact on non-visual text input. 

Looking back on the surveys done by~\citet{azenkot-exploring} on early speech input, we found it interesting that individuals who were BLV rated the accuracy of speech input just under 4 on a 5-point scale. While it is difficult to make direct comparisons between separate participant pools, we found it noteworthy that a decade later, nearly all of our participants had concerns with the accuracy of speech input. One possible explanation is that, while speech input has improved since the interviews by~\citet{azenkot-exploring}, the expectations of users have also grown. Even if that is the case, we feel that improvements to dictation could have a large impact on users with visual impairments.

One possible limitation to this work arises from how we recruited our participants. People who both signed up for the National Federation of the Blind mailing list and were the first to respond to our email advertisement could have more-than-average experience with technology, and may not necessarily be representative of all adults who are blind. Additionally, all our participants had been blind for a minimum of 19 years, which was before the 2007 release of the first iPhone that removed most physical buttons. People who have become blind more recently may have different experiences with non-visual text input stemming from their use of a touchscreen device prior to losing their vision. Further studies are required to determine how these differences in background and experiences might impact participants' views and needs.

\section{Conclusion}
We conducted semi-structured interviews with 12 adults who were legally blind. We found that speech dictation was the most common text input method they used on mobile devices, followed by an onscreen keyboard with a screen reader. The most common themes that we identified across multiple participants' experiences were 1) the poor accuracy of dictation, 2) difficulty entering text in noisy environments, and 3) difficulty correcting errors in entered text. We recommend using these themes as target areas for future research on non-visual text input. 

From the themes we identified, we distilled five suggested directions for future research: 1) improve dictation accuracy, 2) rely less on or enhance audio feedback, 3) improve the error correction process, 4) ensure new input methods have a low barrier to entry, and 5) provide more fluid non-visual word predictions. We hope our themes and future research directions distilled from our interviewees' lived experiences will help guide research to improve the efficiency and usability of non-visual text input.

\balance

\begin{acks}
This work was supported by NSF IIS-1909248 and by NSF Graduate Research Fellowship DGE-2034833. We thank the participants and the National Federation of the Blind for their assistance in recruitment.
\end{acks}

\bibliographystyle{ACM-Reference-Format}
\bibliography{references}

\end{document}